\documentclass[12pt]{article}

\usepackage{mathrsfs}
\usepackage{amsfonts}
\usepackage{tikz}
\usepackage{amssymb,bm}
\usepackage{amsmath}
\usepackage{amsthm}
\usepackage{bbding}
\usetikzlibrary{matrix,shapes,snakes}

\allowdisplaybreaks
\newtheorem{thm}{Theorem}[section]

\newtheorem{lem}[thm]{Lemma}

\newtheorem{rmk}[thm]{Remark}

\def\proof{\vskip 1mm\noindent{\it Proof.}\quad}
\newcommand{\qedd}{\hspace*{\fill}$\Box$\medskip}

\def\deg{\hbox{\rm{deg\,}}}

\begin{document}

\title{A new proof to complexity of dual basis of
 a type I optimal normal basis\thanks{Partially
supported by National Basic Research Program of China
(2011CB302400).}}

\author{Baofeng Wu\thanks{Key Laboratory of Mathematics
Mechanization, AMSS, Chinese Academy of Sciences,
 Beijing 100190,  China. Email: wubaofeng@amss.ac.cn},
 Kai Zhou\thanks{Key Laboratory of Mathematics
Mechanization, AMSS, Chinese Academy of Sciences,
 Beijing 100190,  China. Email: kzhou@amss.ac.cn}, Zhuojun Liu\thanks{Key Laboratory of Mathematics
Mechanization, AMSS, Chinese Academy of Sciences,
 Beijing 100190,  China. Email: zliu@mmrc.iss.ac.cn}}
 \date{}

\maketitle

\begin{abstract}
The complexity of dual basis of a type I optimal normal basis of
$\mathbb{F}_{q^n}$ over $\mathbb{F}_{q}$ was determined to be $3n-3$
or $3n-2$ according as $q$ is even or odd, respectively, by Z.-X.
Wan and K. Zhou in 2007. We give a new proof to this result by
clearly deriving the dual of a type I optimal normal basis with the
aid of a lemma on the dual of a polynomial basis.\vskip .5em

\end{abstract}


\section{Introduction}
Let $\mathbb{F}_{q^n}$ be an extension of the finite field
$\mathbb{F}_{q}$. A basis of the form $\{\alpha, \alpha^q, \ldots,
\alpha^{q^{n-1}}\}$ is called a normal basis, and $\alpha$ is called
a normal basis generator in this case. Normal bases have great
advantages against the common polynomial bases in arithmetic of
finite fields. A simple property about a normal basis is that it has
normal dual basis. By dual basis $\{\beta_1^*,\ldots, \beta_n^*\}$
of a given basis $\{\beta_1,\ldots, \beta_n\}$, we mean
$\mathrm{Tr}_{\mathbb{F}_{q^n}/\mathbb{F}_{q}}(\beta_i\beta_j^*)=\delta_{ij}$,
$1\leq i, j\leq n$.

For a normal basis generator $\alpha$, denote
$\alpha_i=\alpha^{q^i}$, $i=0,\ldots,n-1$, and
$N=\{\alpha_0,\ldots,\alpha_{n-1}\}$. Let
\begin{equation}\label{multab}
\alpha\cdot\alpha_{i}=\sum_{j=0}^{n-1}t_{ij}\alpha_j,~~t_{ij}\in\mathbb{F}_{q},~0\leq
i\leq n-1.
\end{equation}
 The number of nonzero elements in the $n\times n$ matrix
$T=(t_{ij})$ is called the complexity of the normal basis $N$. We
denote it by $C_N$. It can be shown that $C_N\geq 2n-1$, and $N$ is
called optimal when the lower bound is attained.

A type I optimal normal basis of $\mathbb{F}_{q^n}$ over
$\mathbb{F}_{q}$ is the normal basis formed by the $n$ nonunit
$(n+1)$-th roots of unity, where $n+1$ is a prime and $q$ is
primitive modulo $n+1$ \cite{liao,mene}. In \cite{zxwan} the dual of
a type I optimal normal basis was studied, and its complexity was
determined to be $3n-3$ or $3n-2$ according as $q$ is even or odd,
respectively.

Two bases $\{\beta_1,\ldots, \beta_n\}$ and $\{\beta_1',\ldots,
\beta_n'\}$ of $\mathbb{F}_{q^n}$ over $\mathbb{F}_{q}$ are called
equivalent, or weakly equivalent, if $\beta_i=c\beta_i'$,
$i=1,\ldots,n$ for some $c\in \mathbb{F}_{q}$, or
$\mathbb{F}_{q^n}$, respectively. It is easy to see that equivalent
normal bases share the same complexity.

In this paper, we clearly derive the dual $M$ of a type I optimal
normal basis $N$ through a new approach and thus rediscover the
complexity of it. The main auxiliary lemma we use will be presented
in section 2, and in section 3, we will propose our method to get
$M$ and its complexity, which is totally different from that in
\cite{zxwan} and seems more simple and easier to understand.


\section{Auxiliary lemma on the dual of a polynomial basis}
The following lemma, proposed as an exercise in \cite[Chapter
2]{lidl}, is the main result we will need in computing the dual of a
type I optimal normal basis. We will include a short proof for it
here for completeness.

\begin{lem}\label{dualpoly}
Let $\mathbb{F}_{q^n}=\mathbb{F}_{q}(\alpha)$ be an extension of the
finite field $\mathbb{F}_{q}$, and $f(x)\in\mathbb{F}_{q}[x]$ be the
minimal polynomial of $\alpha$. Assume
\[\frac{f(x)}{x-\alpha}=\beta_0+\beta_1x+\cdots+\beta_{n-1}x^{n-1}\in\mathbb{F}_{q^n}[x].\]
Then the dual basis of the polynomial basis
$\{1,\alpha,\ldots,\alpha^{n-1}\}$ is
$\frac{1}{f'(\alpha)}\{\beta_0,\beta_1,\ldots,\beta_{n-1}\}:=\{\frac{\beta_0}{f'(\alpha)},
\frac{\beta_1}{f'(\alpha)},\ldots,\frac{\beta_{n-1}}{f'(\alpha)}\}$,
where $f'$ is the formal derivative of $f$.
\end{lem}

\proof Let $\sigma_i$ be the $i$-th Frobenius automorphism of
$\mathbb{F}_{q^n}/\mathbb{F}_{q}$ and
$\alpha_i=\sigma_i(\alpha)=\alpha^{q^i}$, $0\leq i\leq n-1$. Then
$f(x)=\prod_{i=0}^{n-1}(x-\alpha_i)$. For $l=0,1,\ldots,n-1$, we
construct the auxiliary polynomial
\[F_l(x)=\sum_{i=0}^{n-1}\frac{f(x)}{x-\alpha_i}\frac{\alpha_i^l}{f'(\alpha_i)}-x^l.\]
It is easy to check $F_l(\alpha_i)=0$ for $i=0,1,\ldots,n-1$. Thus
$F_l(x)\equiv 0$ since $\deg F_l(x)\leq n-1$. On the other hand,

\begin{eqnarray*}
  x^l=x^l+F_l(x) &=&
  \sum_{i=0}^{n-1}\sigma_i\Big(\frac{f(x)}{x-\alpha}\frac{\alpha^l}{f'(\alpha)}\Big)\\
  &=&\sum_{i=0}^{n-1}\sigma_i\Big(\sum_{j=0}^{n-1}\beta_jx^j\frac{\alpha^l}{f'(\alpha)}\Big) \\
   &=&\sum_{j=0}^{n-1}\Bigg(\sum_{i=0}^{n-1}\sigma_i\Big(\beta_j\frac{\alpha^l}{f'(\alpha)}\Big)\Bigg)x^j\\
   &=&\sum_{j=0}^{n-1}\mathrm{Tr}_{\mathbb{F}_{q^n}/\mathbb{F}_{q}}\Big(\frac{\beta_j}{f'(\alpha)}\alpha^l\Big)x^j.
\end{eqnarray*}

\noindent So we get
$\mathrm{Tr}_{\mathbb{F}_{q^n}/\mathbb{F}_{q}}\big(\frac{\beta_j}{f'(\alpha)}\alpha^l\big)=\delta_{jl}$
for any $0\leq j,l\leq n-1$, which is what we want to prove.\qedd

\begin{rmk}
In fact, the lemma and the proof hold for any separable algebraic
extension $K(\alpha)$ of a field $K$ with generator $\alpha$.
\end{rmk}


\section{Dual basis of a type I optimal normal basis and its
complexity}

In the remaining part of the paper, we suppose $n+1$ is a prime and
$q$ is primitive modulo $n+1$. Let $\alpha$ be a nonunit $(n+1)$-th
root of unity. Then the set of all nonunit $(n+1)$-th roots of unity
is $N=\{\alpha, \alpha^2, \ldots, \alpha^n\}$.

\begin{lem}{\cite{mene}}\label{onb} The minimal polynomial of $\alpha$ is
$f(x)=\sum_{i=0}^{n}x^i$. Furthermore,
$N=\{\alpha,\alpha^q,\ldots,\alpha^{q^{n-1}}\}$ and it forms an
optimal normal basis of $\mathbb{F}_{q^n}$ over $\mathbb{F}_{q}$.
\end{lem}

The normal basis $N$ in Lemma \ref{onb} is often called a type I
optimal normal basis. Since $N=\{\alpha, \alpha^2, \ldots,
\alpha^n\}=\alpha\{1, \alpha, \ldots, \alpha^{n-1}\}$, we find it
weakly equivalent to the polynomial basis $\{1, \alpha, \ldots,
\alpha^{n-1}\}$. Thus if we can compute the dual of the polynomial
basis, say
$\frac{1}{f'(\alpha)}\{\beta_0,\beta_1,\ldots,\beta_{n-1}\}$ by
Lemma \ref{dualpoly}, then we can get the dual of $N$ of the form
$M=\frac{1}{\alpha
f'(\alpha)}\{\beta_0,\beta_1,\ldots,\beta_{n-1}\}$. The remaining
task is just to compute $\alpha f'(\alpha)$ and $\beta_0, \ldots,
\beta_{n-1}$.

\begin{lem}\label{task1}
\[\alpha f'(\alpha)=\frac{n+1}{\alpha-1}.\]
\end{lem}

\proof Since $f(x)=\prod_{i=1}^{n}(x-\alpha^i)$, we know that
\[f'(\alpha)=\prod_{i=2}^{n}(\alpha-\alpha^i)=\alpha^{n-1}\prod_{i=1}^{n-1}(1-\alpha^i)=\alpha^{n-1}\frac{f(1)}{1-\alpha^n}=\frac{\alpha^{n-1}}{1-\alpha^n}(n+1).\]
Then \[\alpha
f'(\alpha)=\frac{\alpha^n}{1-\alpha^n}(n+1)=\frac{\alpha^{n+1}}{\alpha-\alpha^{n+1}}(n+1)=\frac{n+1}{\alpha-1}.\]\qedd

\begin{lem}\label{task2}
\[\beta_i=\frac{\alpha^{n-i}-1}{\alpha-1},~~i=0,1,\ldots,n-1.\]
\end{lem}

\proof Since
$\frac{f(x)}{x-\alpha}=\prod_{i=2}^{n}(x-\alpha^i)=\sum_{i=0}^{n-1}\beta_ix^i$,
we get $\beta_{n-1}=1$ and
$\beta_{n-i}=(-1)^{i-1}s_{i-1}(\alpha^2,\alpha^3,\ldots,\alpha^n)$
for $i=2,\ldots, n$, where $s_k$ stands for the $k$-th elementary
symmetric polynomial.

As $f(x)=\prod_{i=1}^{n}(x-\alpha^i)=\sum_{i=0}^{n}x^i$, we know
that
\begin{eqnarray*}
  1 &=& (-1)^{i}s_{i}(\alpha,\alpha^2,\alpha^3,\ldots,\alpha^n)\\
  &=& (-1)^{i}[s_{i}(\alpha^2,\alpha^3,\ldots,\alpha^n)+\alpha s_{i-1}(\alpha^2,\alpha^3,\ldots,\alpha^n)] \\
   &=&(-1)^{i}s_{i}(\alpha^2,\alpha^3,\ldots,\alpha^n)-\alpha(-1)^{i-1}s_{i-1}(\alpha^2,\alpha^3,\ldots,\alpha^n)\\
   &=&\beta_{n-i-1}-\alpha\beta_{n-i}.
\end{eqnarray*}
\noindent Thus
\begin{eqnarray*}
\beta_{n-i-1}&=&1+\alpha\beta_{n-i}\\
&=&1+\alpha(1+\alpha\beta_{n-i+1})\\
&=&1+\alpha+\alpha^2\beta_{n-i+1} \\
&=&\cdots\\
&=&1+\alpha+\alpha^2+\cdots+\alpha^i\beta_{n-1}\\
&=&1+\alpha+\cdots+\alpha^i\\
&=&\frac{\alpha^{i+1}-1}{\alpha-1}
\end{eqnarray*}
\noindent for $i=1,\ldots, n-1$. So
$\beta_i=\frac{\alpha^{n-i}-1}{\alpha-1}$ for
$i=0,1,\ldots,n-1$.\qedd

From Lemma \ref{task1} and Lemma \ref{task2}, we can clearly get
that the dual of $N$ is
\[M=\frac{1}{\alpha
f'(\alpha)}\{\beta_0,\beta_1,\ldots,\beta_{n-1}\}=\frac{1}{n+1}\{\alpha^n-1,
\alpha^{n-1}-1,\ldots,\alpha-1\}.\] The normal basis $M$ can be
generated by $\tilde{\alpha}=\frac{\alpha-1}{n+1}$, the minimal
polynomial of which is
\begin{eqnarray*}
  \tilde{f}(x) &=& \prod_{i=1}^n (x-\frac{\alpha^i-1}{n+1}) \\
   &=& \frac{1}{(n+1)^n}\prod_{i=1}^n ((n+1)x+1-\alpha^i) \\
  &=& \frac{1}{(n+1)^n}f((n+1)x+1) \\
   &=&\frac{1}{(n+1)^n}\sum_{i=0}^{n}((n+1)x+1)^i\\
   &=&\frac{1}{(n+1)^n}\sum_{i=0}^{n}\sum_{j=0}^{i}{i\choose
   j}(n+1)^jx^j\\
   &=&\sum_{j=0}^{n}\bigg(\sum_{i=j}^{n}\frac{{i\choose
   j}}{(n+1)^{n-j}}\bigg)x^j.
\end{eqnarray*}

\begin{thm}\label{comple}
The complexity of the dual of a type I optimal normal basis is
$3n-3$ when $q$ is even, and $3n-2$ when $q$ is odd.
\end{thm}

\proof We use the notations above. We only need to compute the
complexity of $(n+1)M=\{\alpha^n-1,
\alpha^{n-1}-1,\ldots,\alpha-1\}$ as it is equivalent to $M$.

Since \begin{eqnarray*}
        (\alpha-1)(\alpha^i-1) &=& \alpha^{i+1}-\alpha-\alpha^i+1 \\
         &=&(\alpha^{i+1}-1)-(\alpha^i-1)-(\alpha-1) \\
        &=& \left\{\begin{array}{cl}
        (\alpha^2-1)-2(\alpha-1)&\text{if}~ i=1\\
        (\alpha^{i+1}-1)-(\alpha^i-1)-(\alpha-1)&\text{if} ~2\leq
        i\leq n-1\\
        -(\alpha^{n}-1)-(\alpha-1)&\text{if}~ i=n,
        \end{array}\right.
      \end{eqnarray*}
\noindent we can get that the complexity of $(n+1)M$ is
$2+3(n-2)+1=3n-3$ when $q$ is even, and is $2+3(n-2)+2=3n-2$ when
$q$ is odd.\qedd

The matrix $T$ determined by (\ref{multab}) for a normal basis is
often called its multiplication table. For the type I optimal normal
basis $N=\{\alpha, \alpha^q, \ldots, \alpha^{q^{n-1}}\}$, its
multiplication table is of the form

\[T_\alpha=P\left(\begin{matrix}
0&1&&\\
&0&\ddots&\\
&&\ddots&1\\
-1&-1&\cdots&-1
\end{matrix}\right) P^{-1},\]
where $P$ is an $n\times n$ permutation matrix with
\[\left(\begin{matrix}
q^0\\q^1\\\vdots\\q^{n-1}
\end{matrix}\right)\equiv P
\left(\begin{matrix} 1\\2\\\vdots\\n
\end{matrix}\right)~\mod\;(n+1).
\]
(More properties on this matrix $P$ can be found in \cite{liao}.)
From the proof of Theorem \ref{comple}, we actually obtain that the
multiplication table of the dual basis $M=\{\tilde{\alpha},
\tilde{\alpha}^q, \ldots, \tilde{\alpha}^{q^{n-1}}\}$ is of the form
\[T_{\tilde{\alpha}}=\frac{1}{n+1}P\left(\begin{matrix}
-2&1&&&&\\
-1&-1&1&&&\\
\vdots&&\ddots&\ddots&&\\
\vdots&&&\ddots&\ddots&\\
-1&&&&-1&1\\
-1&&&&&-1
\end{matrix}\right) P^{-1}.\]

\begin{rmk}
Note that the main observation we make in deriving the dual of a
type I optimal normal basis generated by $\alpha$ is the weak
equivalence between it and the polynomial basis generated by
$\alpha$, so it is a natural question that when will a normal basis
generator $\beta$ generate a normal basis weakly equivalent to the
polynomial basis generated by it in a general finite field
$\mathbb{F}_{q^n}$. We remark that this happens if and only if
$\beta$ is a type I optimal normal basis generator. The reason is
simple: if there exists some $\gamma\in\mathbb{F}_{q^n}$ such that
$N=\{\beta, \beta^q, \ldots, \beta^{q^{n-1}}\}=\gamma\{1, \beta,
\ldots, \beta^{n-1}\}$, we know that $\beta=\gamma\beta^k$ for some
$k\in\mathbb{Z}$, $0\leq k\leq n-1$. Thus $\gamma=\beta^{1-k}$. If
$k\neq 0$, $\gamma\beta^{k-1}=\beta^0=1$, which is impossible since
it is an element of $N$. So $\gamma=\beta$. A normal basis of the
form $\{\beta, \beta^2, \ldots, \beta^n\}$ must be a type I optimal
normal basis, as its complexity is no more than $2n-1$.
\end{rmk}


\end{document}